# Lunar Eclipse Phenomena:

# Modeled and Explained

Anthony Mallama

anthony.mallama@gmail.com

2022 July 21

Abstract

A model based on celestial geometry and atmospheric physics predicts the dimming and the color of lunar eclipses. Corresponding visual magnitudes and color indices for eclipses from year 2000 through 2050 are listed. The enlargement of the Earth's umbral shadow reported by observers for over 300 years is explained. The geometrical aspects of the model are the sizes and separations of the Sun, Moon and Earth. Atmospheric effects include refraction, absorption and focusing of sunlight.

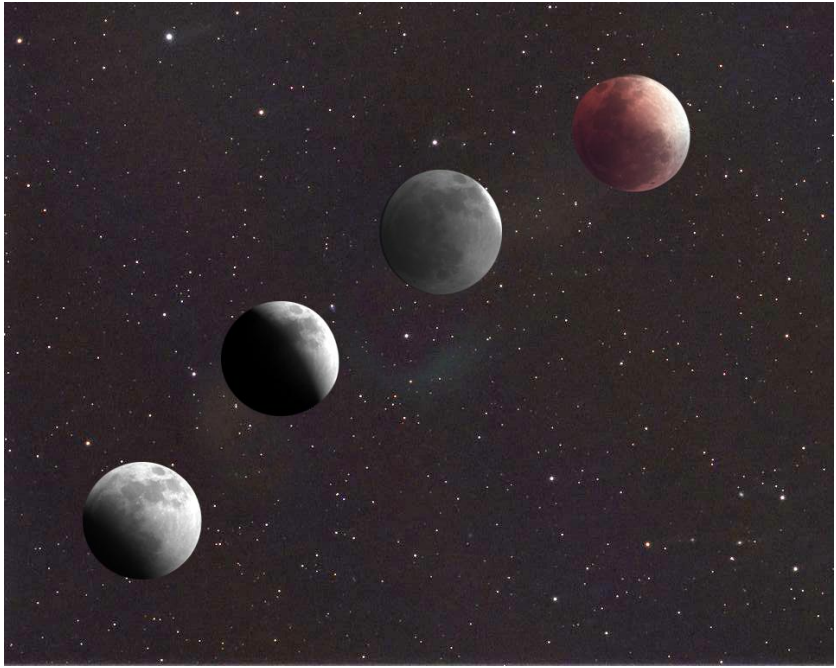

The phases of a lunar eclipse are illustrated from deep penumbral shading on the lower left through totality on the upper right. Robert L Beers Jr. recorded the lunar images with an ASI2600MM camera and a Redcat51 telescope during the eclipse of May 15/16, 2002. He then added them to a star field photographed earlier with a DSLR camera.

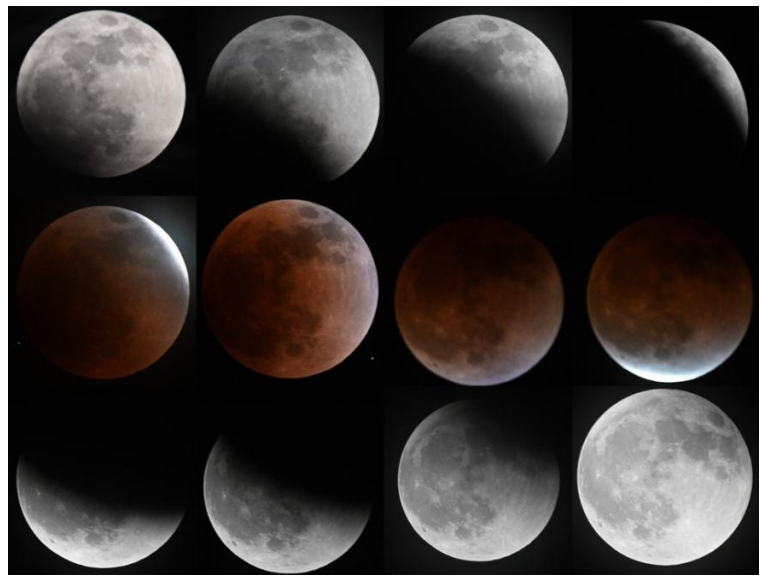

Chuck Story obtained these images of the same eclipse using a Nikon Z6 camera with a 500 mm lens set at f/5.6.

Both of the astrophotographers who made these images are members of the Chagrin Valley Astronomical Society.

## 1. Introduction

Lunar eclipses are eye-catching astronomical displays accessible to millions of people. Some of the phenomena associated with these events are simple to understand: the Earth is blocking sunlight so the Moon grows faint. Others are less obvious: the Earth's atmosphere causes the Moon to turn a reddish color. Still others are complex and have been an enigma for centuries: the Earth's shadow appears to be larger than expected.

This last phenomenon, the perceived enlargement of the umbra, is particularly interesting. As recently as the lunar eclipse of November 2021, amateur astronomers were being asked to help measure it (Sinnott, 2021). This enlargement is explained here.

Section 2 of this paper illustrates the basic geometry of a lunar eclipse. Section 3 shows how sunlight on its way toward the eclipsed Moon interacts with the Earth's atmosphere. The principal atmospheric effects are refraction, absorption and focusing. Section 4 describes a lunar eclipse model that combines geometrical and atmospheric effects. Section 5 explains the brightness and color of a lunar eclipse while Section 6 addresses the umbral enlargement.

## 2. Eclipse geometry

Lunar eclipses happen during the full phase of the Moon when it is opposite the Sun, but not every full moon produces an eclipse. The plane of the Earth's orbit around the Sun, the ecliptic, lies in the plane of the screen (or paper) in Figure 2.1. Meanwhile the Moon's orbit around the Earth is inclined to the ecliptic, and the two planes intersect along the line of nodes. The lunar orbit in the Figure is depicted by a solid line above the ecliptic and a dashed line below.

A lunar eclipse occurs when the full moon lies along the line of nodes. Thus, in the upper part of the Figure, the Moon (marked by letter *M*) at position *A* is full but not eclipsed because it is not on the line of nodes. Likewise, at position *B* it is on the line of nodes but it is not eclipsed because it is not full. However, in the lower part of the Figure, at position *C*, the Moon is eclipsed because it is full and on the line of nodes. Eclipses of the Moon occur about twice a year when the full moon is near one of its nodes.

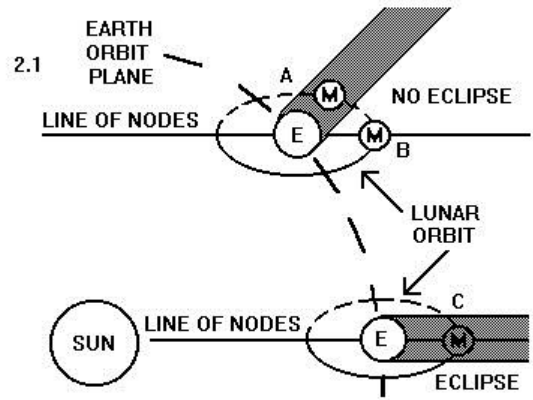

The half-degree diameter of the solar disk at the Earth and Moon distance produces the distinction between the penumbra and umbra of the terrestrial shadow. In the penumbral shadow part of the Sun is visible behind the Earth's limb, while throughout the umbra all the Sun is geometrically blocked by the Earth, as shown in Figure 2.2. Solar illumination in the penumbra shades from full intensity at its outer boundary to deep shadow at the umbral boundary.

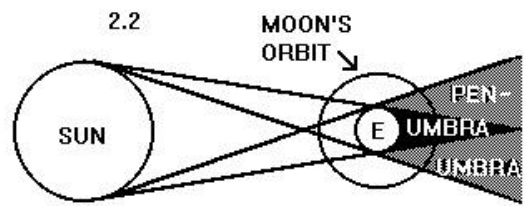

There are three phases of a lunar eclipse. In the penumbral phase, some or all of the Moon is in the penumbra but no part is in the umbra; in the partial phase some of the Moon is in the umbra; and in the total phase, all of the Moon in the umbra. A total eclipse has total, partial, and penumbral phases; a partial eclipse has a partial and a penumbral phase; and a penumbral eclipse only has a penumbral phase.

The sizes of the shadow regions are determined by the geometry of the Sun, Earth, and Moon. Umbral and penumbral dimensions are compared to the Moon's diameter in Figure 2.3. The Moon is approximately equal in size to the width of the penumbral annulus, while a little less than three lunar widths span the umbra. The conditions of contact between the leading and trailing limbs of the Moon and the umbra are labeled U1 for the first contact, U2 for the second contact. U1 and U4 indicate the beginning and end of the partial phase, while totality lasts from U2 until U3. The labels P1 and P4 mark the first and last penumbral contacts.

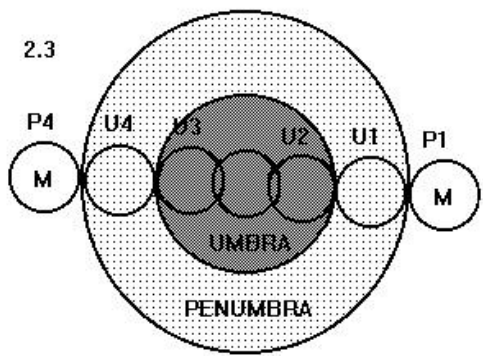

The geometrical factors cited above predict eclipse durations that are too short when compared to actual timings of U1 and U4 shadow contacts. So, most almanacs base their calculations on a hypothetical umbra that is about 2% larger than its geometrical size.

This apparent enlargement was first reported in 1687 by Phillipe de la Hire (Westfall and Sheehan, 2015). Some astronomers postulated an actual 'absorbing layer' high in the terrestrial atmosphere to account for shadow enlargement (Link, 1963). However atmospheric scientists never detected any such absorbing layer, even with sensitive meteorological instruments.

The actual cause of the shadow enlargement phenomena is now understood, and it will be explained in Section 6 of this paper. In preparation for understanding this enlargement and other eclipse phenomena, the next section explains the behavior of sunlight in the Earth's atmosphere.

## 3. Terrestrial atmospheric effects

If the Earth was airless, geometry would render the Moon invisible during the total phase of an eclipse. However the Moon remains visible during totality, due to the Earth's atmosphere. The three principal effects of the atmosphere, refraction, absorption and focusing are discussed in turn.

### 3.1 Refraction

The Sun appears to be displaced by about 35 arc-minutes (35') above its actual position when it is seen on the horizon. Since a ray of light that just grazes sea level on its way to an eclipsed Moon travels twice as far through the atmosphere as the setting Sun ray, it is refracted by 70'. This tangential ray is deflected by about 7800 km at the lunar distance, an amount that is larger than the radius of the Earth. That deflection accounts for the sunlight at the center of the shadow and it renders the Moon visible even when centered in the umbra.

If the Moon were a lot closer to the Earth it would pass through a portion of the shadow inside of the cone produced by grazing rays from around the terrestrial limb. Then it would completely disappear from view during a central eclipse. See the schematic in Figure 3.1.

Refraction in the atmosphere increases progressively with depth. So, light that bends into the shadow is diluted in strength. The intensity depends on the degree of refraction a light ray has experienced, and on the distance it travels from above the Earth's limb to the point where it is measured.

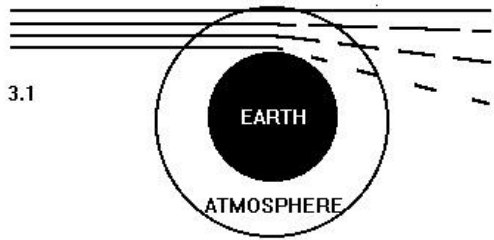

The relationship between refraction and intensity was used by astronomers decades ago to infer the densities of planetary atmospheres. These studies were based on observations of the gradual dimming of stars as they approached planets. (See Wasserman and Veverka 1973 and Hunten and Veverka 1976, for examples.) This relationship can be inverted to model lunar eclipses because the state of the Earth's atmosphere is already understood.

Atmospheric density, refraction and light intensity interact as follows. The amount of refraction of a light ray in a planetary atmosphere is proportional to the density of the air it traverses. The density of the atmosphere at 50 km altitude is only about 1/1000 of its density at the sea level. Since the surface ray bends almost 8000 km (as noted above), the proportionality rule indicates that refraction of the 50 km ray is about 8 km.

Furthermore, the rate of change of density with altitude determines the *atmospheric scale height*. This quantity is about 8 km in the low to middle parts of the Earth's atmosphere. So, at 50 km the deviation of a light ray due to refraction is equal to the scale height. This unique condition identifies the ray that is reduced to exactly half its original intensity, and it serves as the starting location for computing the brightness of refracted light.

The distance from the shadow center where this half-intensity ray passes at the lunar distance is the value *d(0)* in Table 3.1. Intensity versus deflection is computed from Equation 1

$$((i(0) / i(r)) - 2) + \ln((i(0) / i(r)) - 1) = d(0) - d$$

Equation 1

where *i(0)* is the full intensity of light, *i(r)* is the intensity after refraction at the lunar distance (so *i(r)/i(0)* is relative intensity), *d(0)* is the distance measured in scale heights from the center of the Earth's shadow to the starting point referred to above, and *d* is the distance measured in scale heights from the shadow center to the point in question. For more information related to the Table see Appendix A.

------

Table 3.1  Refraction of a point source of light

| Relative Intensity | d-d(0) scale ht | d-d(0) km | d km | Minimum Altitude |
|---|---|---|---|---|
| 1.00 | infinity | | | |
| 0.75 | +1.8 | +15 | 6435 | 64 |
| 0.50 | 0 | 0 | 6420 | 50 |
| 0.25 | -3.1 | -25 | 6395 | 37 |
| 0.10 | -10 | -82 | 6338 | 32 |
| 0.03 | -35 | -278 | 6142 | 25 |
| 0.01 | -102 | -816 | 5604 | 18 |
| 0.003 | -337 | -2696 | 3724 | 8 |
| 0.001 | -1005 | -8040 | -1620 | 0 |

------

The calculations in the Table do not account for the light ray's wavelength because refraction is practically independent of color. This fact is apparent from the appearance of the setting Sun. While the refracted solar image is displaced upwards by 35', very little if any color dispersion is seen. Sometimes a tiny fleck of green can be seen for a few seconds after the rest of the Sun's image disappears below the horizon. However, even this *green flash* phenomenon only represents about 1' of color dispersion.

## 3.2 Absorption and transmission

While refraction is nearly independent of color, absorption introduces hue into the lunar eclipse. Rayleigh scattering due to molecular absorption produces the blue color of the sky and the redness of the setting Sun. These colors occur because blue light is scattered more than red light. The total absorption and transmission of light passing through the atmosphere to the Earth's surface has been measured many times by astronomers performing stellar photometry. Typical values of transmission range from about 73% for a vertical ray of blue light up to 90% for red. The vertical ray encounters one unit *air mass*.

Transmission for a tangential ray passing through any minimum altitude on its course to the eclipsed Moon can be determined by tracing its path. Altitude is computed at small intervals along the way and atmospheric density is determined at each point. The cumulative air mass for the ray is compared to the unit air mass encountered by the vertical ray. Finally the resulting transmission for the tangential rays is determined from Equation 2,

$$i(t) / i(0) = E^m$$

Equation 2

where *i(t)* is the intensity of light after passage through the atmosphere, *i(0)* is the intensity of light before passage (so, *i(t)/i(0)* is the relative intensity), *E* is the value of transmission for a unit air mass discussed above, and *m* is the cumulative air mass that the ray of light has traversed compared to that for a vertical ray. Representative values are listed in Table 3.2.

## 3.3 Focusing

In additional to refraction and absorption, light passing tangentially at any altitude above the Earth's surface is condensed into a smaller radius within the shadow. This geometric focusing of light enhances its intensity by the inverse of distance from the central axis of the shadow.

Table 3.2  Transmission of green light

```
Relative                    Minimum
Intensity      m            altitude

1.0           0.0            ---
0.9           0.6             32
0.8           1.3             27
0.7           1.8             25
0.6           2.7             22
0.5           3.8             20
0.4           5.0             18
0.3           6.5             17
0.2           8.5             15
0.1           13              13
0.01          24               7.9
0.001         37               4.8
0.0001        50               2.6
0.00001       62               0.8
```

## 3.4 Summary of atmospheric effects

A ray of light undergoes refraction, absorption, and focusing in the terrestrial atmosphere. The intensity upon reaching the lunar distance is the product of these three effects. This intensity may become small but it never quite falls to zero in the Earth's shadow. The next section describes an eclipse model that combines geometric and atmospheric effects.

## 4. Lunar eclipse model

The geometric and atmospheric effects described in the two preceding sections can be joined to produce a detailed model for lunar eclipses. Such a model predicts the brightness and color of the eclipsed Moon. This section describes how the model works.

The refraction, absorption, and focusing effects for a point source of light are computed. Then the brightness in the shadow from the center out to the edge of the penumbra is calculated from the product of the three atmospheric effects and by factoring in the apparent size of the solar disk. Finally, the integrated brightness of the Moon is determined by combining a Moon-sized disk with the distribution of brightness in the shadow.

Numerical results for disk resolved and integrated lunar brightness, as a function of position in the shadow, are listed in Appendix B.

The disk resolved magnitudes from Table B-1 are plotted in Figure 4.1 Full brightness corresponding to the outer boundary of the penumbra is indicated at the left edge of the Figure. The magnitude scale on that axis can be used to gauge the decrease of brightness in the shadow, noting that 5 magnitudes corresponds to a factor of 100 in intensity. Going rightward toward the shadow center, brightness is seen to decline slowly at first. Then a very rapid fading accompanied by separation of the curve into three color-dependent branches commences about a third of the way in.

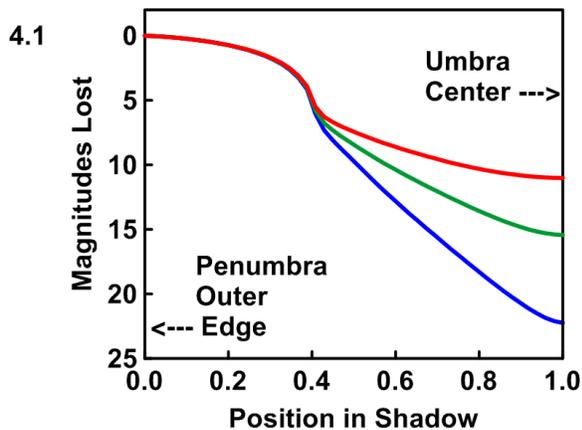

The accelerated decline in brightness corresponds to the transition from penumbra to umbra. Fading in blue light is the most rapid, and that of red light the slowest, as a result of color dependent absorption.

The intensity of blue light in the center of the umbra has fallen by 22 magnitudes, which is almost a billion in intensity. Red light is down by 11 magnitudes, or 20,000 times. The vast difference between these intensities explains the deep color on the Moon when seen in the umbra.

The integrated brightness and color of the Moon is calculated by assigning its position in the shadow, and then summing the elements of brightness on its surface according to their distances from the shadow center as in Table B-1. The resulting integrated magnitudes are given in Table B-2.

The integrated magnitudes at mid-eclipse times for actual events may be computed by determining the Moon's position in the shadow and using the data from Table B-2. Appendix C lists those magnitudes along with color indices for eclipses occurring between years 2000 and 2050.

## 5. Validation of the model and explanation of eclipse brightness and color

The validity of the lunar eclipse modeling described above can be evaluated by comparing observed magnitudes to calculated values.

The model indicates that visual observers should see only a modest fading of the Moon's brightness as it goes from magnitude -12.7 outside of eclipse to -12.0 at the beginning of the partial phase. A steeper decline to magnitude -10.0 will be seen when the center of the Moon crosses the umbra boundary. More dramatic dimming follows until the Moon drops to magnitude -3.5 at the onset of totality. During eclipses when the Moon reaches the center of the umbra it will fade even more to magnitude +1.4. The brightness ratio between magnitudes -12.7 and +1.4 is nearly one-half million.

The model light curve for integrated visual magnitude is indicated by the line in Figure 5.1. Penumbral contacts P1 and P4 correspond to the left end of the horizontal axis. The right end of the axis represents a Moon located centrally in the umbra. Observations plotted in black in the Figure were taken from Sky and Telescope

Magazine. Some magnitudes were obtained by J. Westfall during the eclipses of February 1971 and January 1972.

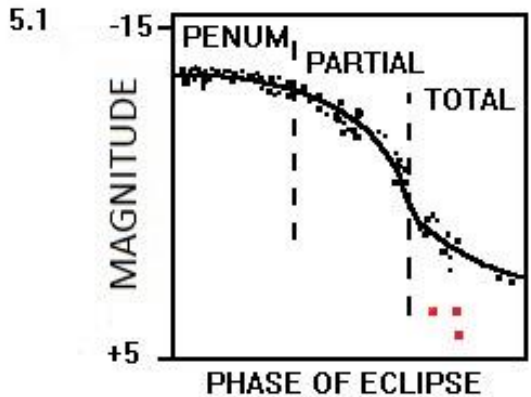

The close fit between the model and observations validates the brightness aspect of model. The modest discrepancies are consistent with observational error, as well as model uncertainties that result from neglecting the lunar albedo features.

The Moon's integrated brightness in the umbra can occasionally be significantly fainter than the model predicts. The orange symbols in the lower right of the Figure are for abnormally dark eclipses. These followed major volcanic eruptions which injected great quantities of aerosols into the stratosphere. Those particles absorb light that would be refracted into the umbra under clear atmosphere conditions. Thus the model indicates an approximate upper limit to the lunar brightness during totality.

## 6. Explanation of the umbral enlargement

Having accounted for the dimming and the color of the eclipsed Moon, the other important phenomenon remaining to be discussed is the perceived enlargement of the Earth's umbra.

This increase can be understood by plotting intensity as a function of position. That was done in Figure 4.1 and in this section Figure 6.1 shows the same relationship at an expanded scale in the region where the brightness change is most rapid. Horizontal axis values correspond to distance outside of a straight line from the solar limb to the terrestrial limb projected to the Moon. This geometric umbra boundary corresponds to zero on that axis.

Without refraction, the light curve would plummet to infinite magnitude (zero brightness) at the geometric boundary. However refraction begins to reduce the slope of the line at a higher altitude. The bottom portion of the Figure shows the rate of brightness change, which is steepest at 211 km outside of that boundary. Visual observers will perceive the umbra boundary somewhere near the radius of steepest brightness gradient. So, the eclipse model successfully reproduces the shadow enlargement phenomenon at least qualitatively.

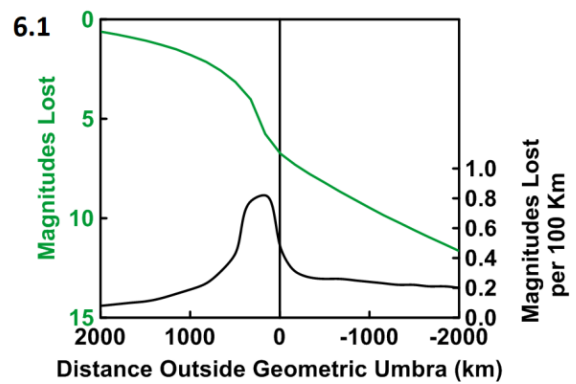

The amount of enlargement according to the model is larger than that determined by observations though. Herald and Sinnott (2014) determined a value of 87 km based on visual data.

The actual degree of enlargement will depend on the state of the atmosphere. The modeled value is for a perfectly clear atmosphere and any extra attenuation will reduce it. Absorption counteracts the effect of refraction (which is what enlarges the umbra) and thus absorption makes the umbra appear less enlarged One source of added absorption is ozone and in particular the Chappuis absorption band in the visible part of the spectrum. The model does not take ozone into account. Additionally, clouds and high terrain along the Earth's limb block sunlight at low altitudes where it is most strongly refracted. These effects will combine to reduce the umbral enlargement from 211 km to a smaller value.

Visual perception may also explain the smaller enlargement that is derived empirically. Observers are instructed to record the time that *umbra's edge* covers a lunar crater or other feature. That edge may be perceived to be towards the darker side of the peak in Figure 6.1 where the landmark would be very dim. That instruction is different than asking observers to record when the *steepest gradient* covers a feature. The gradient would correspond to the model enlargement but the observers may be recording something slightly different.

## 7. Conclusions

A lunar eclipse model that combines celestial geometry with terrestrial atmospheric effects is described. The atmosphere causes refraction, absorption and focusing of light that impacts the eclipsed Moon. The model successfully explains the observed brightness and color of the eclipse phenomenon as well as the enlargement of the umbra.

## Appendix A – Additional notes for Table 3.1

For the half-intensity ray and from Equation 1 in Section 3, i(0)/i(r) is 2 and d - d(0) = 0. Additionally, i(r)/i(0) near the center of the shadow is about 1/1000.

The minimum height of the half-intensity ray can be derived from Equation A-1

$$\theta = H / D = v * \sqrt{2 * \pi * a / H}$$

<div style="text-align: right;">Equation A-1</div>

where θ is the bending angle, *H* is the scale height, *D* is distance to the Moon, v is index of refraction of the gas minus one, and *a* is the Earth's radius. The height is directly related to v. For example, v is 3E-7 for the Earth-Moon, while v at sea level is 3E-4, indicating that the ray's minimum height is at 0.001 of sea level pressure, or 50 km altitude.

## Appendix B – Magnitudes in the shadow

The modeling described in Section 4 predicts the following magnitudes of brightness in the Earth's shadow. The colors correspond to B (blue), V (green), R (red), and I (near-infrared) filters of the Johnson-Cousins UBVRI photometric system. They are computed for a clear atmosphere, otherwise the brightness may be reduced. Distances of the Sun and Moon are for an eclipse were the geometric penumbral annulus width equals the lunar diameter. For other cases of solar and lunar distances the *Pos'n* scale should be adjusted accordingly.

Table B-1 lists magnitudes of intensity lost at a given point in the shadow for the 'disk resolved' case. Table B-2 lists magnitudes of the whole lunar disk for the 'disk integrated' case.

Table B.1 The intensity of blue, green, red, and infrared light in the shadow of the Earth is listed in magnitudes lost. The *Shadow Pos'n* ranges from 0.000 at the outer penumbral boundary to 1.000 at the center of the shadow.

```
Shadow     ---- Magnitudes Lost ----
Pos'n      Blue    Green    Red     IR

0.000      0.00    0.00     0.00    0.00
0.020      0.02    0.02     0.03    0.03
0.041      0.06    0.06     0.07    0.07
0.061      0.11    0.11     0.12    0.13
0.082      0.17    0.17     0.18    0.19
0.102      0.24    0.24     0.24    0.25
0.122      0.32    0.32     0.32    0.33
0.143      0.41    0.41     0.41    0.42
0.163      0.51    0.51     0.51    0.52
0.184      0.63    0.63     0.63    0.62
0.204      0.76    0.76     0.75    0.74
```

```
0.224    0.92    0.91    0.90    0.88
0.245    1.09    1.08    1.06    1.04
0.265    1.30    1.28    1.26    1.22
0.286    1.54    1.51    1.49    1.43
0.306    1.83    1.80    1.77    1.70
0.327    2.19    2.14    2.10    2.01
0.347    2.64    2.58    2.53    2.41
0.367    3.24    3.15    3.08    2.93
0.388    4.15    4.02    3.91    3.71
0.408    6.04    5.76    5.50    5.19
0.429    7.32    6.76    6.28    5.82
0.449    8.07    7.31    6.68    6.11
0.469    8.74    7.78    7.01    6.33
0.490    9.39    8.22    7.30    6.52
0.510   10.04    8.65    7.57    6.68
0.531   10.71    9.07    7.83    6.83
0.551   11.35    9.47    8.08    6.97
0.571   11.97    9.86    8.31    7.09
0.592   12.58   10.23    8.52    7.21
0.612   13.17   10.59    8.74    7.32
0.632   13.76   10.94    8.94    7.42
0.653   14.33   11.29    9.13    7.51
0.673   14.90   11.63    9.32    7.59
0.694   15.46   11.96    9.50    7.67
0.714   16.02   12.28    9.68    7.74
0.734   16.57   12.60    9.85    7.80
0.755   17.11   12.91   10.00    7.85
0.775   17.65   13.22   10.15    7.89
0.796   18.19   13.52   10.29    7.91
0.816   18.72   13.80   10.42    7.93
0.837   19.24   14.08   10.53    7.94
0.857   19.74   14.34   10.64    7.96
0.878   20.23   14.58   10.73    7.97
0.898   20.68   14.80   10.81    7.99
0.918   21.12   15.00   10.88    8.00
0.939   21.52   15.17   10.94    8.01
0.959   21.86   15.30   10.98    8.01
0.980   22.11   15.39   11.01    8.02
1.000   22.24   15.44   11.02    8.02
```

Table B.2  The integrated apparent magnitude of the eclipsed Moon is tabulated for blue, green, red, and near-infrared light. The *Moon Pos'n* ranges from 0.000 at first and last penumbral contact, to 1.000 when the Moon is exactly centered in the umbra.

```
Moon         Integrated Magnitude
Pos'n    Blue   Green    Red     IR
0.000   -11.82 -12.73 -13.65 -14.04
0.016   -11.81 -12.72 -13.64 -14.03
0.033   -11.81 -12.72 -13.64 -14.03
0.050   -11.81 -12.72 -13.64 -14.03
0.066   -11.81 -12.72 -13.64 -14.03
0.083   -11.80 -12.71 -13.63 -14.02
0.100   -11.80 -12.71 -13.62 -14.01
0.116   -11.79 -12.69 -13.61 -14.00
0.133   -11.77 -12.68 -13.60 -13.99
0.150   -11.75 -12.66 -13.58 -13.97
0.166   -11.73 -12.64 -13.56 -13.95
0.183   -11.71 -12.62 -13.54 -13.92
0.200   -11.68 -12.59 -13.51 -13.89
0.216   -11.64 -12.55 -13.47 -13.86
0.233   -11.60 -12.51 -13.43 -13.82
0.250   -11.55 -12.46 -13.38 -13.77
0.266   -11.50 -12.41 -13.33 -13.72
0.283   -11.43 -12.34 -13.26 -13.65
0.300   -11.36 -12.27 -13.19 -13.59
0.316   -11.28 -12.19 -13.11 -13.51
0.333   -11.19 -12.10 -13.02 -13.42
0.350   -11.09 -12.00 -12.92 -13.32
0.366   -10.97 -11.89 -12.81 -13.21
0.383   -10.85 -11.76 -12.69 -13.09
0.400   -10.71 -11.63 -12.56 -12.96
0.416   -10.57 -11.48 -12.41 -12.82
0.433   -10.41 -11.33 -12.26 -12.67
0.450   -10.23 -11.15 -12.09 -12.50
0.466   -10.04 -10.96 -11.90 -12.32
0.483    -9.83 -10.76 -11.70 -12.12
0.500    -9.60 -10.53 -11.47 -11.91
0.516    -9.34 -10.28 -11.23 -11.67
0.533    -9.06 -10.00 -10.96 -11.41
0.550    -8.74  -9.69 -10.65 -11.12
0.566    -8.39  -9.34 -10.31 -10.80
0.583    -7.98  -8.94  -9.92 -10.43
0.600    -7.50  -8.48  -9.48 -10.01
0.616    -6.94  -7.94  -8.95  -9.53
0.633    -6.25  -7.28  -8.34  -8.97
0.650    -5.40  -6.48  -7.61  -8.35
0.666    -4.30  -5.49  -6.78  -7.72
0.683    -2.91  -4.40  -6.02  -7.22
0.700    -1.85  -3.71  -5.59  -6.99
0.716    -1.13  -3.24  -5.32  -6.85
0.733    -0.49  -2.83  -5.07  -6.74
0.750     0.14  -2.43  -4.85  -6.64
0.766     0.76  -2.05  -4.64  -6.55
0.783     1.38  -1.68  -4.44  -6.47
0.800     1.98  -1.33  -4.26  -6.40
0.816     2.56  -1.00  -4.09  -6.34
0.833     3.13  -0.67  -3.93  -6.28
0.850     3.68  -0.36  -3.78  -6.24
0.866     4.21  -0.07  -3.64  -6.20
0.883     4.74   0.21  -3.51  -6.16
0.900     5.25   0.47  -3.40  -6.14
0.916     5.73   0.72  -3.30  -6.12
0.933     6.19   0.94  -3.22  -6.10
0.950     6.61   1.13  -3.15  -6.09
0.966     6.97   1.28  -3.10  -6.08
0.983     7.25   1.39  -3.06  -6.07
1.000     7.39   1.44  -3.05  -6.07
```

**Appendix C – Eclipses from year 2000 through 2050 with magnitudes and colors**

T = total, P = partial and N = penumbral
CI = color index (degree of redness)

```
    - UT Date -   Type   Mag    CI
    21 JAN 2000    T     -1     3.5
    16 JUL 2000    T     +1     5
     9 JAN 2001    T     -2     3
```

```
 5 JUL 2001   P   -10     1
30 DEC 2001   N   -12     1
26 MAY 2002   N   -12.5   1
24 JUN 2002   N   -12.5   1
20 NOV 2002   N   -12.5   1
16 MAY 2003   T    -2.5   2.5
 9 NOV 2003   T    -3.5   2
 4 MAY 2004   T    -1     3.5
28 OCT 2004   T    -1     3.5
24 APR 2005   N   -12     1
17 OCT 2005   P   -11.5   1
14 MAR 2006   N   -12     1
 7 SEP 2006   P   -11.5   1
 3 MAR 2007   T    -1.5   3
28 AUG 2007   T     0     4.5
21 FEB 2008   T    -2.5   3
16 AUG 2008   P    -6     1
 9 FEB 2009   N   -12     1
 7 JUL 2009   N   -12.5   1
 6 AUG 2009   N   -12.5   1
31 DEC 2009   P   -11.5   1
26 JUN 2010   P   -10     1
21 DEC 2010   T    -1.5   3
15 JUN 2011   T    +1     5
10 DEC 2011   T    -2.5   3
 4 JUN 2012   P   -10.5   1
28 NOV 2012   N   -12.5   1
25 APR 2013   P   -12     1
25 MAY 2013   N   -12.5   1
18 OCT 2013   N   -12.5   1
15 APR 2014   T    -1     3.5
 8 OCT 2014   T    -2     3
 4 APR 2015   T    -3.5   2
28 SEP 2015   T    -1.5   3
23 MAR 2016   N   -12.5   1
18 AUG 2016   N   -12.5   1
16 SEP 2016   N   -12     1
11 FEB 2017   N   -12     1
 7 AUG 2017   P   -11     1
31 JAN 2018   T    -1     3.5
27 JUL 2018   T    +0.5   5
21 JAN 2019   T    -2     3
16 JUL 2019   P    -9     1
10 JAN 2020   N   -12     1
 5 JUN 2020   N   -12.5   1
30 NOV 2020   N   -12.5   1
26 MAY 2021   T    -3.5   2
19 NOV 2021   P    -3.5   2
16 MAY 2022   T    -0.5   4
 8 NOV 2022   T    -0.5   4
 5 MAY 2023   N   -12     1
28 OCT 2023   P   -11.5   1
25 MAR 2024   N   -12     1
18 SEP 2024   P   -11.5   1
14 MAR 2025   T    -2     3
 7 SEP 2025   T    -0.5   4
 3 MAR 2026   T    -2     3
28 AUG 2026   P    -4     1.5
20 FEB 2027   N   -12     1
18 JUL 2027   N   -12.5   1
17 AUG 2027   N   -12.5   1
12 JAN 2028   P   -11.5   1
 6 JUL 2028   P   -10.5   1
31 DEC 2028   T    -1.5   3
26 JUN 2029   T    +1.5   5.5
20 DEC 2029   T    -2.5   3
15 JUN 2030   P   -10     1
 9 DEC 2030   N   -12     1
 7 MAY 2031   N   -12     1
 5 JUN 2031   N   -12.5   1
30 OCT 2031   N   -12.5   1
25 APR 2032   T    -2     3
18 OCT 2032   T    -2.5   3
14 APR 2033   T    -2.5   3
 8 OCT 2033   T    -1     3.5
 3 APR 2034   N   -12.5   1
28 SEP 2034   P   -12     1
22 FEB 2035   N   -12     1
19 AUG 2035   P   -11.5   1
11 Feb 2036   T    -1.5   3.5
 7 Aug 2036   T    -0.5   4
31 Jan 2037   T    -2     3
27 Jul 2037   P    -7     1
21 Jan 2038   N   -12.5   1
16 Jul 2038   N   -12.5   1
17 Jun 2038   N   -12.5   1
11 Dec 2038   N   -12.5   1
 6 Jun 2039   P    -7.5   1
30 Nov 2039   P    -4     1.5
26 May 2040   T     0     4.5
18 Nov 2040   T    -0.5   4
16 May 2041   P   -12     1
 8 Nov 2041   P   -11.5   1
 5 Apr 2042   N   -12     1
29 Sep 2042   P   -12     1
25 Mar 2043   T    -2.5   2.5
19 Sep 2043   T    -2     3
13 Mar 2044   T    -2.5   3
 7 Sep 2044   T    -3.5   2
 3 Mar 2045   N   -12     1
27 Aug 2045   N   -12.5   1
22 Jan 2046   P   -12     1
18 Jul 2046   P   -11.5   1
12 Jan 2047   T    -1.5   3
 7 Jul 2047   T    +1.5   5.5
 1 Jan 2048   T    -2.5   2.5
26 Jun 2048   P    -9.5   1
20 Dec 2048   N   -12     1
17 May 2049   N   -12.5   1
15 Jun 2049   N   -12.5   1
 9 Nov 2049   N   -12.5   1
 6 May 2050   T    -3     2.5
30 Oct 2050   T    -3.5   2
```

### End notes

Some of the material in this paper was derived from a work written by the author in 1996. That more extensive study, titled *Eclipses, Atmospheres and Global Change,* examined eclipses of all the satellites of the solar system, in addition to the relationship between atmospheric absorption and climate change. It was not published in a journal, nor was it widely distributed to colleagues.